\documentclass[preprintnumbers, floatfix, letterpaper, nofootinbib, twocolumn]{revtex4}
\pdfoutput=1
\usepackage{graphicx}
\usepackage{microtype}
\usepackage{amsmath,amsthm}
\usepackage{amssymb}
\usepackage{subfigure}
\usepackage{hyperref}
\usepackage{url}
\usepackage{xcolor}
\usepackage{color}
\usepackage{mathrsfs}
\usepackage{calrsfs}
\usepackage{amsfonts}
\usepackage{latexsym}
\usepackage{ragged2e}
\usepackage{epsfig}
\usepackage{textcomp}
\usepackage{phaistos}
\usepackage{lipsum}

\makeatletter
\renewcommand\@makefnmark{\hbox{\@textsuperscript{\normalfont\@thefnmark}}}
\renewcommand\@makefntext[1]{%
  \parindent 1em\noindent
            \hb@xt@1.8em{%
                \hss\@textsuperscript{\normalfont\@thefnmark}}#1}
\makeatother

\usepackage{caption}
\usepackage{placeins}
\DeclareCaptionJustification{justified}{\leftskip=0pt \rightskip=0pt \parfillskip=0pt plus 1fil}
\definecolor{palatinateblue}{rgb}{0.15, 0.23, 0.89}
\definecolor{brightpink}{rgb}{1.0, 0.0, 0.5}

\hypersetup{ linktoc=all,
    colorlinks, linkcolor={palatinateblue},
    citecolor={brightpink}, urlcolor={black}
}
\graphicspath{{Images/}}

\graphicspath{{Images/}}



\def\sideremark#1{\ifvmode\leavevmode\fi\vadjust{\vbox to0pt{\vss
 \hbox to 0pt{\hskip\hsize\hskip1em
 \vbox{\hsize1.5cm\tiny\raggedright\pretolerance10000
 \noindent #1\hfill}\hss}\vbox to8pt{\vfil}\vss}}}%

\begin{document}

\title{Spectral Localization in Cavity-Mediated Entanglement Harvesting}
\author{Hao Xu}
\thanks{Corresponding author}
\email{haoxu@yzu.edu.cn}
\affiliation{Center for Gravitation and Cosmology, College of Physical Science and Technology, Yangzhou University, \\180 Siwangting Road, Yangzhou City, Jiangsu Province 225002, China}

\begin{abstract}
We investigate the relation between spectral localization and entanglement harvesting in an analytically solvable model of two qubits coupled to a leaky single-mode cavity, which in turn couples to a continuous electromagnetic bath. We derive the no-jump concurrence at a prescribed reference gate time in closed form, $\mathcal C_g(Q)=2e^{-\pi/(2Q)}(1+e^{-\pi/(2Q)})/(1+3e^{-\pi/Q})$, where $Q\equiv|\Delta|/\kappa$ is the ratio of the qubit-cavity detuning $\Delta$ to the cavity linewidth $\kappa$. This reference time is inherited from the first maximally entangling unitary gate. At finite loss, $\mathcal C_g$ is the conditional concurrence at this prescribed readout time. In the high-$Q$ limit, $\mathcal C_g\simeq1-\pi^2/(16Q^2)$, while in the low-$Q$ limit it decays exponentially. For the Lorentzian spectral family, $Q$ is proportional to a detuning-normalized inverse participation ratio and simultaneously measures the balance between coherent exchange and collective decay. This yields a compact, model-specific description of the crossover from nearly unit conditional entanglement to overdamped suppression. The predicted $\mathcal C_g(Q)$ curve can, in principle, be examined in circuit QED with ideal monitoring and no-jump post-selection.
\end{abstract}

\maketitle

\section{Introduction}

Entanglement harvesting stands as a central protocol in relativistic quantum information~\cite{Reznik2005,PozasKerstjens2015,Salton2015,Tjoa2021,Xu:2026zyg}. Two initially uncorrelated, spatially localized quantum detectors, typically Unruh-DeWitt (UDW) two-level systems~\cite{Unruh1976,DeWitt1979,Hodgkinson2012,MartinMartinez2013,Hummer2016,Xu:2023tdt,Xu:2025smb}, acquire quantum entanglement solely through local coupling to a common quantum field. This entanglement emerges even when the detectors remain spacelike separated throughout the interaction, thereby precluding any direct signal exchange. The underlying mechanism does not rely on direct detector-detector couplings; rather, it exploits the nonlocal correlations inherent to the field vacuum. The quantum vacuum is a complex, structured state governed by fluctuations with correlations spanning all spacetime regions, which is a profound consequence of the Reeh--Schlieder theorem~\cite{Reeh1961,Haag1996,Witten2018} and its corollaries~\cite{Summers1985,Summers1987}. The detectors convert these pre-existing field correlations into entanglement between their internal degrees of freedom~\cite{Reznik2003}. Valentini~\cite{Valentini1991} first showed that vacuum correlations can be extracted by localized probes; since then, this protocol has been extensively generalized to both flat and curved spacetimes~\cite{PozasKerstjens2015,LinHu2010,Ng2018,PozasKerstjens2017,MartinMartinez2016}.

From a quantum information-theoretic standpoint, this mechanism is fundamentally bounded by the quantum data processing inequality: the entanglement obtained by the detectors cannot surpass the pre-existing correlations within the field across the two coupling regions~\cite{Schumacher1996}. The detectors do not generate entanglement, and they merely harvest it from the vacuum reservoir. Within the standard UDW framework, where pointlike detectors are linearly coupled to a massless scalar field in Minkowski spacetime, the harvested entanglement remains perturbatively weak. The concurrence~\cite{Wootters1998} vanishes in the weak-coupling limit and, for generic parameters, falls well below the threshold requisite for Bell inequality violations~\cite{PozasKerstjens2015}. This suppression is inherent to the perturbative regime. Vacuum correlations are ultraviolet (UV) divergent and formally unbounded, but local probes can access only a tiny fraction of them. The local coupling acts as a stringent low-pass filter, allowing only a small portion of the vacuum's nonlocal correlations to be transferred to the detectors.

This limitation naturally motivates a question: under what conditions can two localized detectors extract maximal entanglement (i.e., a Bell state with unit concurrence) from a quantum field within finite time? Moreover, how does this extraction efficiency depend on the field’s spectral structure? Beyond its foundational implications for quantum field theory, achieving unit-concurrence Bell states represents a benchmark for entanglement distribution and quantum communication networks~\cite{Kimble2008}.

For the standard continuous scalar-field protocols considered in the literature, no regime of unit concurrence at finite time has been established. The unit concurrence of the reduced two-detector state would necessitate a pure maximally entangled state, implying that the joint detector-field state must factorize at that instant~\cite{Wootters1998}. In generic regimes, nonvanishing detector-field coupling entangles the detectors with the field. Because the two detectors couple to field degrees of freedom at distinct spatial points, the resulting detector-field correlations generally prevent exact refactorization in the standard setups considered here. Moreover, a continuous spectrum typically disperses detector information over infinitely many modes, so an exact finite-time revival is not expected without additional structure or fine tuning. Consequently, the reduced two-detector state is generally mixed, so unit concurrence is not attained in the standard continuous-field regimes analyzed to date. This picture aligns with structural no-go theorems for entanglement extraction~\cite{Simidzija2018}, which delineate coupling regimes (such as degenerate detectors or single $\delta$-couplings) where continuous-field harvesting fails entirely. It is further corroborated by perturbative and nonperturbative analyses showing that harvested entanglement scales as $\mathcal{O}(\lambda^{2})$~\cite{PozasKerstjens2015} and diminishes beyond an optimal coupling strength~\cite{Simidzija2018}, with no known regime achieving unit concurrence. In the standard continuous-field settings relevant here, the field therefore acts as an effectively irreversible reservoir, making unit concurrence unavailable within the regimes established to date.

In contrast, a qualitatively different scenario arises when the quantum field theory is replaced by a single discrete cavity mode operating in the large-detuning regime of cavity QED~\cite{Zheng2000,Blais2004,Blais2021}. Here, the frequency mismatch between the qubit transitions and the cavity resonance strongly suppresses real energy exchange, leaving the cavity only virtually excited. Through second-order virtual-photon processes, the mode mediates an effective, unitary qubit-qubit exchange interaction. Starting from a suitable separable initial state, this dispersive dynamics can deterministically drive the two qubits into a Bell state at a precise gate time, thereby achieving unit concurrence~\cite{Zheng2000}. This mechanism has been experimentally realized in both atomic cavity QED~\cite{Raimond2001,Osnaghi2001} and superconducting circuits, where the cavity-bus serves as a robust platform for deterministic two-qubit gates~\cite{Wallraff2004,Majer2007,DiCarlo2009}.

In this context, entanglement does not arise solely from harvesting pre-existing vacuum correlations, and it is generated via unitary gate operations. The maximal entanglement should therefore not be interpreted as being extracted directly from the vacuum; rather, it is generated by the effective coherent interaction mediated by the cavity mode. The discrete cavity mode can support coherent information return within this mediated interaction, unlike the effectively irreversible continuum behavior of the standard settings discussed above, albeit at the cost of altering the entanglement-generation mechanism. This shift also dictates a change in the required initial state: while ground-state detectors suffice for conventional vacuum harvesting~\cite{PozasKerstjens2015}, the collective ground state $|00\rangle$ in our effective dynamics is dark and stationary (Sec.~\ref{sec:concurrence}), necessitating an excited, factorized preparation. Deploying this very same initial state in the continuous-field UDW setting still yields concurrence that vanishes in the weak-coupling limit. The present calculation thus describes a cavity-mediated harvesting protocol whose operational mechanism differs from standard spacelike UDW harvesting.

Historically, these two lines of research, namely deterministic gate operations in discrete-spectrum cavity QED and entanglement harvesting in continuous-spectrum free fields, have evolved along largely parallel tracks, separated both by their physical inquiries and their theoretical formalisms. The former typically employs Lindblad master equations and quantum trajectory methods~\cite{Wiseman2009}, whereas the latter relies on Wightman function perturbation theory~\cite{PozasKerstjens2015,Salton2015}. This contrast motivates asking, within a controlled cavity model, how a tunable linewidth changes the balance between coherent exchange and loss.

In this work, we bridge this gap with a minimal, analytically solvable model: two qubits couple to a leaky single-mode cavity, which in turn couples to an external zero-temperature electromagnetic continuum. The cavity linewidth $\kappa$, the rate at which the cavity mode leaks into the continuum, serves as a continuously tunable spectral knob. When the cavity is perfect and the linewidth vanishes ($\kappa\to0$), the qubits see a fully discrete spectrum and the dispersive gate regime is recovered. When the cavity is strongly overdamped ($\kappa\to\infty$), the cavity-mediated coherent channel is destroyed and entanglement generation vanishes. Within this cavity construction, the narrow- and broad-linewidth limits form the two ends of a continuously tunable degree of spectral localization. This setup therefore provides a controlled interpolation between narrow and broad Lorentzian responses while retaining the cavity as the mediator of the qubit-bath coupling.

For arbitrary intermediate linewidths, we derive a closed-form expression for the no-jump concurrence at the reference gate time, $\mathcal C_g(Q)$. It depends on the dimensionless ratio $Q\equiv|\Delta|/\kappa=|J|/\Gamma$, where $\Delta$ is the qubit-cavity detuning, $J$ the coherent exchange, and $\Gamma$ the collective decay coefficient. For the Lorentzian spectral family, the same dependence can be parameterized by a detuning-normalized inverse participation ratio (IPR). We use spectral localization to denote this model-specific concentration of the effective spectral response relative to the detuning.

\begin{figure}
\centering
\includegraphics[width=\columnwidth]{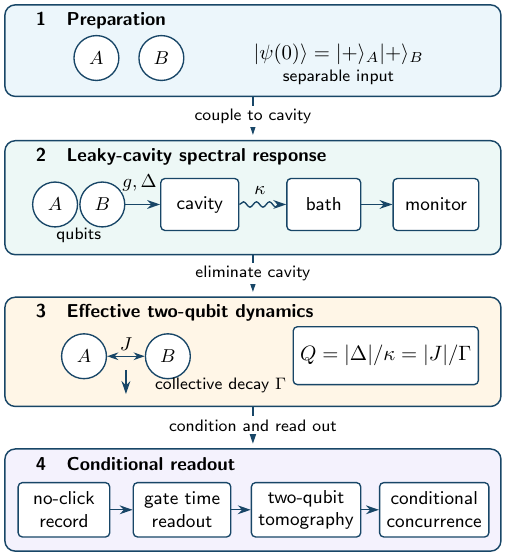}
\caption{Operational sequence of the cavity-mediated protocol. A separable two-qubit input couples to a leaky cavity whose output is monitored. Eliminating the cavity gives coherent exchange $J$ and collective decay $\Gamma$, organized by $Q=|\Delta|/\kappa=|J|/\Gamma$. Selecting a no-click record and performing two-qubit tomography at the reference time yield the conditional concurrence.}\label{fig:workflow}
\end{figure}

The operational sequence is summarized in Fig.~\ref{fig:workflow}. We prepare the separable state $|+\rangle_A|+\rangle_B$, ideally monitor the entire cavity output, and retain runs in which no photon is detected (the no-click, or no-jump, record). After dispersive elimination of the cavity, the selected branch evolves under coherent exchange $J$ and collective decay $\Gamma$. We read out the qubits at $t_g=\pi/(2|J|)$, the first maximally entangling time of the lossless gate. At finite loss, $t_g$ is a prescribed reference time rather than an optimized time. Two-qubit tomography at this time yields $\mathcal C_g$, which is a conditional concurrence because the state is normalized within the selected no-click ensemble.

We emphasize that this model is not a spacetime-resolved simulation of the UDW harvesting protocol. It contains no plane waves, no spacelike separation, and no direct detector-field coupling. Rather, it isolates how spectral width affects the conditional entanglement in the present cavity-mediated setting. Because the present reduced model coarse-grains the spatial structure, the spectral density characterizes the balance between coherent exchange and loss and its effect on the conditional entanglement within this cavity-mediated setting. Accordingly, the single-mode-plus-continuum construction provides a minimal representation of the spectral response relevant to this calculation, in the same spirit as the Jaynes-Cummings model for atom-field interaction.

At a broader conceptual level, the Reeh--Schlieder theorem provides field-theoretic background for the nontrivial correlation structure of the quantum vacuum. The present cavity model does not include two spacelike-separated coupling regions and therefore does not provide a direct quantitative realization of this field-theoretic statement; instead, it isolates the complementary role of spectral structure by showing that the linewidth of a cavity-mediated channel controls the conditional entanglement. For a Lorentzian response, this spectral concentration can be quantified by the detuning-normalized IPR introduced below. This measure invites a formal analogy with Anderson localization, where participation measures quantify the concentration of wave functions, although the underlying physical mechanisms and observables remain distinct in the two settings. The predicted $\mathcal C_g(Q)$ curve is, in principle, accessible in superconducting circuit QED experiments, and its asymptotic behavior provides a test of the present conditional cavity model.

The paper is organized as follows: Sec.~\ref{sec:model} introduces the model; Sec.~\ref{sec:bm} derives the open-system description; Sec.~\ref{sec:effective} eliminates the cavity to obtain the effective qubit dynamics; Sec.~\ref{sec:concurrence} analyzes the conditional concurrence and its spectral interpretation; Sec.~\ref{sec:conclusion} concludes.

\section{Model and Hamiltonian}
\label{sec:model}

We consider two qubits, each modeled as a two-level system with transition frequency $\omega_0$, coupled to a single-mode coplanar waveguide resonator of frequency $\omega_c$. The resonator in turn couples to an external electromagnetic field, which serves as a zero-temperature bath. In the laboratory frame, the total Hamiltonian reads
\begin{equation}
H_{\text{lab}} = H_{q} + H_{c} + H_{qc} + H_{B} + H_{cB},
\label{eq:H_lab}
\end{equation}
where the five terms are defined as follows.

The free qubit Hamiltonian is $H_{q}=\frac{\omega_{0}}{2}(\sigma_{z}^{A}+\sigma_{z}^{B})$, where $\sigma_{z}^{(j)}$ is the Pauli-$Z$ operator of qubit $j=A,B$. The ground state $|0\rangle_{j}$ and the excited state $|1\rangle_{j}$ have energies $-\omega_{0}/2$ and $+\omega_{0}/2$, respectively. For superconducting transmons, typical values are $\omega_{0}/2\pi\sim4{-}8$~GHz~\cite{Koch2007,Blais2021}. The cavity mode is described by $H_{c}=\omega_{c}a^{\dagger}a$, where $a$ ($a^{\dagger}$) is the bosonic annihilation (creation) operator and $[a,a^{\dagger}]=1$.

After applying the rotating-wave approximation (RWA), the qubit-cavity coupling is
\begin{equation}
H_{qc}=g\big(a\sigma_{+}^{A}+a^{\dagger}\sigma_{-}^{A}\big)+g\big(a\sigma_{+}^{B}+a^{\dagger}\sigma_{-}^{B}\big)
\equiv g\big(a\Sigma_{+}+a^{\dagger}\Sigma_{-}\big),
\label{eq:H_qc}
\end{equation}
where $g$ is the coupling strength of each qubit to the cavity, $\sigma_{\pm}^{(j)}=\frac{1}{2}(\sigma_{x}^{(j)}\pm i\sigma_{y}^{(j)})$ are the raising and lowering operators of qubit $j$, and $\Sigma_{\pm}\equiv\sigma_{\pm}^{A}+\sigma_{\pm}^{B}$ are the corresponding collective operators. The RWA discards the counter-rotating terms $a\sigma_{-}$ and $a^{\dagger}\sigma_{+}$; it is well justified because $g/2\pi\sim50{-}100$~MHz is far smaller than $\omega_{0},\omega_{c}/2\pi\sim5$~GHz~\cite{Blais2021}.

The bath Hamiltonian is
\begin{equation}
H_{B}=\sum_{\mathbf{k}\lambda}\omega_{k}b_{\mathbf{k}\lambda}^{\dagger}b_{\mathbf{k}\lambda},
\quad\omega_{k}=c|\mathbf{k}|,
\label{eq:H_B}
\end{equation}
where $b_{\mathbf{k}\lambda}$ annihilates a bath photon with wave vector $\mathbf{k}$ and polarization $\lambda$. The cavity-bath coupling, also within the RWA, is
\begin{equation}
H_{cB}=\sum_{\mathbf{k},\lambda}\big(\eta_{\mathbf{k}\lambda}ab_{\mathbf{k}\lambda}^{\dagger}
+\eta_{\mathbf{k}\lambda}^{*}a^{\dagger}b_{\mathbf{k}\lambda}\big),
\label{eq:H_cB}
\end{equation}
where the constants $\eta_{\mathbf{k}\lambda}$ quantify the spatial overlap between the cavity standing-wave mode and the external plane-wave modes at the cavity mirrors. These constants carry all spatial information, namely the cavity mode function evaluated at the mirror positions and the phase factors of the external plane-wave modes.

\subsection{Cavity linewidth}

Within the Born-Markov framework~\cite{Wiseman2009}, the entire influence of the bath on the cavity mode is captured by the bath spectral density,
\begin{equation}
\mathcal{J}_{\text{bath}}(\omega)\equiv\sum_{\mathbf{k},\lambda}|\eta_{\mathbf{k}\lambda}|^{2}\,\delta(\omega-\omega_{k}).
\label{eq:J_bath}
\end{equation}
By direct analogy with Einstein's $A$ coefficient in atomic physics, the cavity photon-number decay rate $\kappa$ is then given by Fermi's golden rule evaluated at the cavity frequency,
\begin{equation}
\kappa\equiv 2\pi\,\mathcal{J}_{\text{bath}}(\omega_{c})
=2\pi\sum_{\mathbf{k},\lambda}|\eta_{\mathbf{k}\lambda}|^{2}\,\delta(\omega_{c}-\omega_{k}).
\label{eq:kappa}
\end{equation}
Physically, $\kappa$ is proportional to the squared cavity-bath coupling strength times the density of external modes at the cavity frequency. The corresponding cavity-amplitude decay rate is $\kappa/2$. It is the single parameter through which the microscopic details of the bath enter the effective qubit dynamics.

\subsection{Co-rotating frame}

To work in the large-detuning regime, where $\Delta\equiv\omega_{0}-\omega_{c}$ satisfies $|\Delta|\gg g$, we transform both the qubits and the cavity to a frame rotating at the qubit frequency $\omega_{0}$,
\begin{align}
|\psi_{I}(t)\rangle &= U(t)|\psi_{S}(t)\rangle, \notag\\
U(t) &= \exp\!\Big(i\frac{\omega_{0}}{2}\Sigma_{z}t\Big)\otimes\exp\!\big(i\omega_{0}a^{\dagger}a\,t\big).
\label{eq:U_def}
\end{align}
This is a change of basis, not of picture. We remain in the Schr\"{o}dinger picture: the state carries all time dependence, the operators $a$ and $\sigma_{\pm}$ are static, but the Hamiltonian that generates the evolution is modified. From the chain rule applied to $i\partial_{t}|\psi_{S}\rangle=H_{\text{lab}}|\psi_{S}\rangle$,
\begin{equation}
i\partial_{t}|\psi_{I}\rangle
=i\dot{U}U^{\dagger}|\psi_{I}\rangle+UH_{\text{lab}}U^{\dagger}|\psi_{I}\rangle,
\end{equation}
we obtain
\begin{equation}
{H_{\text{tot}}=UH_{\text{lab}}U^{\dagger}+i\dot{U}U^{\dagger}}
=H_{0}+H_{B}+V+H_{SB},
\label{eq:H_tot}
\end{equation}
with $H_{0}=-\Delta a^{\dagger}a$, $V=g(a\Sigma_{+}+a^{\dagger}\Sigma_{-})$, and
\begin{equation}
H_{SB}=a e^{-i\omega_{0}t}B^{\dagger}+a^{\dagger}e^{i\omega_{0}t}B,
\quad B\equiv\sum_{\mathbf{k},\lambda}\eta_{\mathbf{k}\lambda}^{*}b_{\mathbf{k}\lambda}.
\label{eq:HSB}
\end{equation}
The connection term $i\dot{U}U^{\dagger}=-\frac{\omega_{0}}{2}\Sigma_{z}-\omega_{0}a^{\dagger}a$ cancels the free qubit Hamiltonian and shifts the cavity to $(\omega_{c}-\omega_{0})a^{\dagger}a=-\Delta a^{\dagger}a$. Likewise, $U(a\sigma_{+})U^{\dagger}=(ae^{-i\omega_{0}t})(\sigma_{+}e^{i\omega_{0}t})=a\sigma_{+}$, so $V$ is time-independent in this frame, a prerequisite for the Born-Markov treatment below. The fast frequencies $\omega_{0},\omega_{c}$ are thus eliminated, leaving only $\Delta$ in $H_{0}$. The phases in $H_{SB}$ keep the cavity-bath resonance at the physical cavity frequency $\omega_{c}$ [cf.\ Eq.~\eqref{eq:kappa}].

\section{Born-Markov Master Equation}
\label{sec:bm}

We now derive the effective open-system dynamics for the qubit-cavity system. Choosing $H_{\text{ref}}\equiv H_{0}+H_{B}$ as the reference Hamiltonian, we move to the interaction picture, in which the system operators evolve as $\tilde{a}(t)=a e^{i\Delta t}$ while $\tilde{\Sigma}_{\pm}(t)=\Sigma_{\pm}$ stay fixed. The total perturbation in this picture is
\begin{equation}
\tilde{H}_{I}(t)=\tilde{V}(t)+\tilde{H}_{SB}(t),
\label{eq:HI}
\end{equation}
where $\tilde{V}(t)=g(a e^{i\Delta t}\Sigma_{+}+a^{\dagger}e^{-i\Delta t}\Sigma_{-})$ and $\tilde{H}_{SB}(t)=a e^{i(\Delta-\omega_{0})t}\tilde{B}^{\dagger}(t)+a^{\dagger}e^{-i(\Delta-\omega_{0})t}\tilde{B}(t)$.
We integrate the von Neumann equation in the interaction picture formally and apply the standard Born-Markov approximations~\cite{Wiseman2009}. The resulting equation for the reduced system state $\tilde{\rho}(t)=\operatorname{Tr}_{B}[\tilde{\rho}_{\text{tot}}(t)]$ is
\begin{align}
\frac{d\tilde{\rho}}{dt}
&=-i\operatorname{Tr}_{B}\!\big([\tilde{H}_{I}(t),\tilde{\rho}(0)\otimes\rho_{B}]\big) \nonumber \\
&-\int_{0}^{\infty}\!\!ds\,\operatorname{Tr}_{B}\!\big([\tilde{H}_{I}(t),[\tilde{H}_{I}(t-s),\tilde{\rho}(t)\otimes\rho_{B}]]\big).
\label{eq:BM}
\end{align}
The first-order term evaluates to $-i[\tilde{V}(t),\tilde{\rho}(t)]$, which describes the coherent qubit-cavity coupling in the interaction picture. The term involving $\tilde{H}_{SB}$ vanishes because the bath vacuum has zero mean. Inserting $\tilde{H}_{I}=\tilde{V}+\tilde{H}_{SB}$ into the second-order term yields three types of contributions.

\subsection{Classification of second-order contributions}

The nested double commutator expands into four terms: $\mathcal{L}_{VV}$ (from $\tilde{V}\tilde{V}$), the cross terms $\mathcal{L}_{VS}+\mathcal{L}_{SV}$, and $\mathcal{L}_{SS}$ (from $\tilde{H}_{SB}\tilde{H}_{SB}$). The cross terms $\mathcal{L}_{VS}$ and $\mathcal{L}_{SV}$ vanish identically upon tracing over the bath, since each contains exactly one bath operator with vanishing vacuum expectation value. 

The $\mathcal{L}_{SS}$ contribution arises purely from the cavity-bath coupling. Evaluating the bath trace with the vacuum correlation function $C(s)=\langle\tilde{B}(s)\tilde{B}^{\dagger}(0)\rangle=\int_{0}^{\infty}d\omega\,\mathcal{J}_{\text{bath}}(\omega)e^{-i\omega s}$ and using $\tilde{a}(t)=a e^{i\Delta t}$, one obtains
\begin{equation}
\mathcal{L}_{SS}[\tilde{\rho}]=\Gamma(\omega_{c})\big[a\tilde{\rho}a^{\dagger}-a^{\dagger}a\tilde{\rho}\big]
+\Gamma^{*}(\omega_{c})\big[a\tilde{\rho}a^{\dagger}-\tilde{\rho}a^{\dagger}a\big],
\label{eq:LSS}
\end{equation}
with $\Gamma(\omega)\equiv\int_{0}^{\infty}ds\,C(s)e^{i\omega s}$. Applying the Sokhotski-Plemelj theorem to separate the real and imaginary parts, with $\mathcal{P}$ denoting the Cauchy principal value,
\begin{equation}
\Gamma(\omega_{c})=\pi\mathcal{J}_{\text{bath}}(\omega_{c})-i\,\mathcal{P}\!\int_{0}^{\infty}\!\!d\omega'\frac{\mathcal{J}_{\text{bath}}(\omega')}{\omega'-\omega_{c}}
\equiv\frac{\kappa}{2}+i\,\delta\omega_{\text{cav}}.
\label{eq:Gamma_decomp}
\end{equation}
The real part yields the cavity Lindblad dissipator $\kappa\mathcal{D}[a]\tilde{\rho}$, where $\mathcal{D}[a]\rho\equiv a\rho a^{\dagger}-\frac{1}{2}\{a^{\dagger}a,\rho\}$ and $\kappa=2\pi\mathcal{J}_{\text{bath}}(\omega_{c})$ is the cavity linewidth, in agreement with Eq.~\eqref{eq:kappa}. The imaginary part $\delta\omega_{\text{cav}}$ is the cavity Lamb shift~\cite{Blais2021}.

The $\mathcal{L}_{VV}$ contribution originates from the qubit-cavity coupling alone and is of order $g^{2}/\Delta$.  Compared with the linear coupling $g(a\Sigma_{+}+a^{\dagger}\Sigma_{-})$ already retained in the coherent part of the dynamics, this contribution is subleading, smaller by one power of $g/\Delta\ll1$, so we drop it from the joint master equation. Its leading effect, the second-order $XY$ exchange, can be generated by the adiabatic elimination of Sec.~\ref{sec:effective} from the linear coupling alone.

\subsection{System Lindblad equation in the Schr\"{o}dinger picture}

Transforming back to the Schr\"{o}dinger picture with respect to $H_{0}$ and absorbing the cavity Lamb shift $\delta\omega_{\text{cav}}$ into a renormalized detuning $\Delta_{\text{ren}}\equiv\Delta-\delta\omega_{\text{cav}}$, which we hereafter denote simply as $\Delta$ since only the renormalized value is experimentally accessible, we obtain the Lindblad equation for the joint qubit-cavity system:
\begin{equation}
{\frac{d\rho}{dt}=-i\big[H_{S},\rho\big]+\kappa\,\mathcal{D}[a]\rho},\quad
H_{S}=-\Delta a^{\dagger}a+g(a\Sigma_{+}+a^{\dagger}\Sigma_{-}).
\label{eq:lindblad}
\end{equation}
Equation~(\ref{eq:lindblad}) is the starting point for all subsequent analysis. It describes two qubits coherently coupled to a damped cavity mode that leaks into the external continuum at rate $\kappa$. The parameter $\kappa$ continuously tunes the spectral structure seen by the qubits. As $\kappa\to0$, the cavity mode is a perfect discrete resonance, while as $\kappa\to\infty$, the cavity becomes transparent and the qubits effectively couple to a featureless continuum. In the next section, we eliminate the cavity degrees of freedom to obtain a closed effective master equation for the qubits alone.

\section{Effective Qubit Dynamics}
\label{sec:effective}

The Lindblad equation~(\ref{eq:lindblad}) describes the joint evolution of the qubits and the cavity mode. We now eliminate the cavity degrees of freedom to obtain a closed effective master equation for the qubits alone. This procedure is justified by the large separation of characteristic timescales in the dispersive regime.

\subsection{Heisenberg-Langevin equation and adiabatic elimination}

From the Lindblad equation~(\ref{eq:lindblad}), the Heisenberg-Langevin equation for the cavity annihilation operator is obtained by evaluating the commutator with $H_{S}$. Using $[H_{S},a]=\Delta a-g\Sigma_{-}$ and the standard input-output formalism~\cite{Wiseman2009}, we find
\begin{equation}
\frac{da}{dt}=i\Delta a-ig\Sigma_{-}-\frac{\kappa}{2}a+\sqrt{\kappa}\,a_{\text{in}}(t),
\label{eq:HE}
\end{equation}
where the input vacuum noise satisfies $\langle a_{\text{in}}(t)\rangle=0$ and $[a_{\text{in}}(t),a_{\text{in}}^{\dagger}(t')]=\delta(t-t')$. The term $-ig\Sigma_{-}$ represents the driving of the cavity by the qubits, $-\frac{\kappa}{2}a$ the cavity damping, and $\sqrt{\kappa}a_{\text{in}}$ the accompanying quantum fluctuations.

The physical basis for adiabatic elimination lies in the hierarchy of timescales. The cavity mode evolves on a characteristic timescale $T_{\text{cav}}\sim 1/|\Delta|$, set by the large detuning $|\Delta|\gg g$; stronger damping only shortens this timescale further. In contrast, the collective qubit operators $\Sigma_{\pm}$ evolve far more slowly. Their Heisenberg equation $\dot{\Sigma}_{-}=ig\,a\,\Sigma_{z}$ is driven by a term proportional to $g$, with no large frequency prefactor. The shortest timescale for qubit dynamics is thus $T_{\text{bit}}\sim 1/g\gg T_{\text{cav}}$. This separation of timescales, where the cavity responds almost instantaneously to the qubits' state, is the defining feature of the dispersive regime.

In the frequency domain, this separation is sharpened by the bandwidth theorem: a signal varying on the timescale $T_{\text{bit}}$ has spectral width $\Delta\omega\sim 1/T_{\text{bit}}\ll|\Delta|$. Together with the qubit equation of motion, the exact frequency-space solution of Eq.~(\ref{eq:HE}),
\begin{equation}
a(\omega)=\frac{-ig\Sigma_{-}(\omega)+\sqrt{\kappa}a_{\text{in}}(\omega)}{-i(\omega+\Delta)+\kappa/2},
\label{eq:a_omega}
\end{equation}
shows that $\Sigma_{-}(\omega)$ is effectively nonzero only for small $|\omega|$. In this narrow band, $\omega$ is negligible compared to $\Delta$ in the denominator. Setting $\omega\to0$, equivalent to $da/dt=0$ in the time domain, yields the quasi-static response
\begin{equation}
a\simeq\frac{-ig\Sigma_{-}}{-i\Delta+\kappa/2}\equiv\alpha\Sigma_{-},
\quad\alpha=\frac{-ig}{-i\Delta+\kappa/2},
\label{eq:adiabatic}
\end{equation}
where the input-noise contribution, which does not affect the unconditional dynamics, has been dropped. This relation replaces the cavity mode operator by a simple algebraic function of the qubit operators. The correction discarded by neglecting $da/dt$ is of relative order $(g/|\Delta|)^{2}\ll1$~\cite{Francica2009}.

\subsection{Effective qubit master equation}

Substituting Eq.~(\ref{eq:adiabatic}) into the system Lindblad equation~(\ref{eq:lindblad}) and tracing over the cavity Hilbert space eliminates all cavity operators. The coherent part of the dynamics arises from two sources. The direct qubit-cavity coupling $V=g(a\Sigma_{+}+a^{\dagger}\Sigma_{-})$ generates qubit-qubit terms under the substitution $a\to\alpha\Sigma_{-}$, and the free cavity Hamiltonian $-\Delta a^{\dagger}a$ contributes, through the same substitution, to both single-qubit Stark shifts and qubit-qubit exchange. The dissipative part originates from the cavity Lindblad term $\kappa\mathcal{D}[a]\rho$, which becomes a collective qubit decay channel after the replacement~\cite{Dicke1954,Gross1982,Ficek2002}.

Collecting the qubit-qubit exchange terms and absorbing all Stark shifts into a renormalization of the qubit frequency, which is the experimentally measured value, we obtain the effective qubit master equation
\begin{equation}
{\frac{d\rho_{q}}{dt}=-i\big[J(\sigma_{+}^{A}\sigma_{-}^{B}+\sigma_{-}^{A}\sigma_{+}^{B}),\rho_{q}\big]
+\Gamma\,\mathcal{D}[\Sigma_{-}]\rho_{q}},
\label{eq:eff_master}
\end{equation}
where $\rho_{q}=\operatorname{Tr}_{\text{cav}}[\rho]$, $\Sigma_{-}=\sigma_{-}^{A}+\sigma_{-}^{B}$, and the two $\kappa$-dependent coefficients are
\begin{equation}
{J(\kappa)=\frac{g^{2}\Delta}{\Delta^{2}+\kappa^{2}/4}},\quad
{\Gamma(\kappa)=\frac{g^{2}\kappa}{\Delta^{2}+\kappa^{2}/4}}.
\label{eq:J_Gamma}
\end{equation}

Equation~(\ref{eq:eff_master}) is the central result of our microscopic derivation. It describes two qubits coupled by a coherent $XY$ exchange of strength $|J(\kappa)|$, which generates entanglement, and subject to collective decay at amplitude rate $\Gamma(\kappa)$, which degrades it. All microscopic details of the original Hamiltonian, namely the cavity frequency, the qubit-cavity coupling, and the bath spectral density, are compressed into these two parameters, which depend solely on the experimentally accessible quantities $g$, $\Delta$, and $\kappa$.

Several features of Eq.~(\ref{eq:J_Gamma}) are worth emphasizing. First, $J$ and $\Gamma$ share the same denominator $\Delta^{2}+\kappa^{2}/4$, the Lorentzian response of the damped cavity: both arise from second-order processes in the qubit-cavity coupling $g$ and inherit the cavity's spectral profile. Second, their ratio $\Gamma/|J|=\kappa/|\Delta|$ is controlled by the single dimensionless parameter $Q\equiv|\Delta|/\kappa$, a detuning-to-linewidth ratio distinct from the conventional cavity quality factor $\omega_c/\kappa$. Third, these limits illustrate the narrow- and broad-linewidth regimes of the present model. When $\kappa\to0$ ($Q\to\infty$), the qubits see a perfectly discrete spectrum: $\Gamma\to0$ and the dynamics is purely unitary, with $J\to g^{2}/\Delta$. When $\kappa\to\infty$ ($Q\to0$), $J\to0$ and entanglement generation is shut off; both rates vanish in this limit, so the qubits asymptotically decouple from the field. In the next section, we compute the conditional concurrence at a specified reference time as a function of $Q$.

\section{Conditional Concurrence at a Reference Gate Time}
\label{sec:concurrence}

The effective master equation~\eqref{eq:eff_master} describes the unconditional evolution of the qubits, i.e., the average over all photodetection records. We use the quantum trajectory formalism~\cite{Wiseman2009}, assume ideal monitoring of the entire cavity output, and retain records with no detected photon. The concurrence below is therefore conditional on the no-jump record; it is not the concurrence of the unconditional density matrix.

\subsection{No-jump evolution and non-Hermitian Hamiltonian}

The Lindblad master equation~(\ref{eq:eff_master}) can be rewritten as
\begin{equation}
\frac{d\rho_{q}}{dt}=-i[H_{\text{int}},\rho_{q}]+\Gamma\Sigma_{-}\rho_{q}\Sigma_{+}
-\frac{\Gamma}{2}\{\Sigma_{+}\Sigma_{-},\rho_{q}\}.
\label{eq:master_split}
\end{equation}
The first dissipative term, $\Gamma\Sigma_{-}\rho_{q}\Sigma_{+}$, describes quantum jumps. When a photon leaks from the cavity into the bath and is detected, the qubit state is projected as $|\psi\rangle\to\Sigma_{-}|\psi\rangle/\|\Sigma_{-}|\psi\rangle\|$. The second term, $-\frac{\Gamma}{2}\{\Sigma_{+}\Sigma_{-},\rho_{q}\}$, describes the smooth, non-unitary evolution conditioned on \emph{no} photon being detected. This no-jump dynamics is generated by the non-Hermitian Hamiltonian
\begin{equation}
H_{\text{nH}}=H_{\text{int}}-i\frac{\Gamma}{2}\Sigma_{+}\Sigma_{-},\quad
H_{\text{int}}\equiv J(\sigma_{+}^{A}\sigma_{-}^{B}+\sigma_{-}^{A}\sigma_{+}^{B}).
\label{eq:HnH}
\end{equation}
For a pure initial state, the no-jump trajectory remains pure at all times and evolves according to $|\tilde{\psi}(t)\rangle=e^{-iH_{\text{nH}}t}|\psi(0)\rangle$, with the squared norm $\langle\tilde{\psi}|\tilde{\psi}\rangle$ giving the probability that no jump has occurred up to time $t$. A weighted average of the no-jump and jump branches rigorously recovers the full Lindblad equation~(\ref{eq:eff_master}).

The non-Hermitian Hamiltonian $H_{\mathrm{nH}}$ is exactly diagonal in the basis $\{|11\rangle,|\Psi_+\rangle,|\Psi_-\rangle,|00\rangle\}$, where $|\Psi_\pm\rangle\equiv(|01\rangle\pm|10\rangle)/\sqrt{2}$ with $|01\rangle\equiv|0\rangle_A|1\rangle_B$, and the complete spectrum reads
\begin{align}
H_{\text{nH}}|11\rangle&=-i\Gamma|11\rangle,\nonumber\\
H_{\text{nH}}|\Psi_{+}\rangle&=(J-i\Gamma)|\Psi_{+}\rangle,\nonumber\\
H_{\text{nH}}|\Psi_{-}\rangle&=-J|\Psi_{-}\rangle,\nonumber\\
H_{\text{nH}}|00\rangle&=0.
\label{eq:HnH_spectrum}
\end{align}
The doubly excited state $|11\rangle$ and the symmetric state $|\Psi_{+}\rangle$ are superradiant. Their unnormalized amplitudes both decay at rate $\Gamma$ in the no-jump branch, while $|\Psi_{+}\rangle$ also accumulates the coherent phase $e^{-iJt}$ under the $XY$ exchange. The antisymmetric state $|\Psi_{-}\rangle$ and the ground state $|00\rangle$ are dark with respect to dissipation. The state $|\Psi_{-}\rangle$ carries the coherent eigenvalue $-J$ and hence acquires a phase $e^{+iJt}$ under the exchange interaction, whereas $|00\rangle$ is strictly stationary. This superradiant/subradiant structure is a direct consequence of the collective nature of the jump operator $\Sigma_{-}$~\cite{Dicke1954,Gross1982,Ficek2002}.

\subsection{Concurrence from the separable initial state $|+\rangle|+\rangle$}

We initialize the qubits in the separable state $|\psi(0)\rangle=|+\rangle_{A}|+\rangle_{B}$, where $|+\rangle=(|0\rangle+|1\rangle)/\sqrt{2}$ is the single-qubit superposition state, distinct from the Bell state $|\Psi_+\rangle$ above. This choice deserves comment in the context of the harvesting framework. In the conventional UDW protocol, detectors are initialized in their ground states and become entangled by exchanging vacuum correlations. In our setup, the collective ground state $|00\rangle$ is a dark state of the collective jump operator $\Sigma_{-}$ and a stationary state of the effective master equation~(\ref{eq:eff_master}). It evolves trivially and can never become entangled, regardless of the cavity parameters. An excited factorized preparation is therefore required, and we choose $|+\rangle|+\rangle$ as a simple balanced separable state. We do not optimize over all separable inputs. This choice is not what makes unit concurrence reachable in the unitary limit. Expanding $|\psi(0)\rangle$ in the eigenbasis of $H_{\mathrm{nH}}$ and applying the no-jump evolution, the normalized conditional state at time $t$ is
\begin{equation}
|\psi(t)\rangle=\frac{\Bigl(e^{-\Gamma t}|11\rangle+e^{-iJt}e^{-\Gamma t}(|10\rangle+|01\rangle)+|00\rangle\Bigr)}{\sqrt{1+3e^{-2\Gamma t}}}.
\label{eq:psi_t}
\end{equation}
The ground-state component $|00\rangle$ is immune to decay, while the singly and doubly excited components decay at the common rate $\Gamma$. The phase factor $e^{-iJt}$ reflects the coherent $XY$ exchange.

For a pure two-qubit state $|\psi\rangle=a|11\rangle+b|10\rangle+c|01\rangle+d|00\rangle$, the Wootters concurrence~\cite{Wootters1998} is $\mathcal{C}=2|ad-bc|$. Inserting the coefficients of Eq.~(\ref{eq:psi_t}) yields
\begin{align}
\mathcal{C}(t)
&= \frac{2e^{-\Gamma t}\bigl|1-e^{-\Gamma t}e^{-2iJt}\bigr|}{1+3e^{-2\Gamma t}} \nonumber \\
&= \frac{2e^{-\Gamma t}\sqrt{1-2e^{-\Gamma t}\cos(2Jt)+e^{-2\Gamma t}}}{1+3e^{-2\Gamma t}}.
\label{eq:Ct}
\end{align}
In the unitary limit $\Gamma\to0$, this reduces to $\mathcal{C}(t)=|\sin(Jt)|$, confirming that the coherent $XY$ interaction alone can generate maximal entanglement.

We use the reference gate time $t_g=\pi/(2|J|)$, inherited from the first maximally entangling time in the unitary limit. At finite loss it fixes the coherent phase but does not maximize $\mathcal C(t)$. Evaluating Eq.~\eqref{eq:Ct} at $t=t_g$ gives
\begin{equation}
\mathcal C_g(Q)=\frac{2e^{-\pi/(2Q)}\bigl(1+e^{-\pi/(2Q)}\bigr)}{1+3e^{-\pi/Q}},\quad
Q\equiv\frac{|J|}{\Gamma}=\frac{|\Delta|}{\kappa}.
\label{eq:Cg}
\end{equation}
Equation~\eqref{eq:Cg} characterizes the fixed initial state, no-jump record, and reference time used here. The maximization over evolution times or input states is a separate problem. The absence of $g$ from Eq.~\eqref{eq:Cg} follows because both $J$ and $\Gamma$ scale as $g^2$, while $t_g$ scales as $g^{-2}$; the physical protocol duration therefore retains its $g$ dependence.

In the high-$Q$ regime, expanding Eq.~\eqref{eq:Cg} gives
\begin{equation}
\mathcal C_g(Q)=1-\frac{\pi^2}{16Q^2}-\frac{\pi^3}{64Q^3}+\mathcal O(Q^{-4}),
\label{eq:highQ}
\end{equation}
showing that the reference-time concurrence is insensitive to weak cavity loss. For $Q=10$, $\mathcal C_g\simeq0.9934$. The leading coefficient $\pi^2/16$ is parameter independent within this model. In the opposite, low-$Q$ regime, the reference-time concurrence decays exponentially,
\begin{equation}
\mathcal C_g(Q)\simeq2e^{-\pi/(2Q)}\xrightarrow{Q\to0}0,
\label{eq:lowQ}
\end{equation}
This suppression concerns readout at $t_g$ in the selected no-jump branch. Other readout times, initial states, and continuous-field protocols define distinct optimization problems. Throughout the range of $Q$, $\mathcal C_g(Q)$ varies smoothly between its high- and low-$Q$ limits for the protocol considered here.

The reference-time no-jump concurrence depends on $g$, $\Delta$, and $\kappa$ only through $Q=|\Delta|/\kappa$. Neither the absolute coupling scale nor the bare frequencies appear independently in Eq.~\eqref{eq:Cg}. Within the present Lorentzian cavity model, this ratio is proportional to the detuning-normalized inverse participation ratio of the effective spectral density, as shown below.

\subsection{Spectral localization and participation ratio}

The microscopic origin of the result~\eqref{eq:Cg} is a single chain of coarse graining. The constants $\eta_{\mathbf{k}\lambda}$ quantifying the cavity-bath interaction enter only through the rate $\kappa=2\pi\sum_{\mathbf{k},\lambda}|\eta_{\mathbf{k}\lambda}|^{2}\delta(\omega_{c}-\omega_{k})$ (Fermi's golden rule~\cite{Blais2021}). The cavity thus acts as a spectral filter, and the qubits, which do not couple directly to the bath, see the effective spectral density
\begin{equation}
\mathcal{J}_{\text{eff}}(\omega)=\frac{g^{2}}{\pi}\frac{\kappa/2}{(\omega-\Delta)^{2}+(\kappa/2)^{2}},
\label{eq:Jeff}
\end{equation}
a Lorentzian of width $\kappa$ centered at the detuning $\Delta$. For this Lorentzian family, with the integrals taken over the full relative-frequency axis, we define the detuning-normalized inverse participation ratio~\cite{Anderson1958,Evers2008,Wegner1980,Mirlin2000,Kramer1993},
\begin{equation}
\mathrm{IPR}_{\Delta}\equiv2|\Delta|\frac{\int_{-\infty}^{\infty} d\omega[\mathcal J_{\mathrm{eff}}(\omega)]^2}{\bigl(\int_{-\infty}^{\infty} d\omega\mathcal J_{\mathrm{eff}}(\omega)\bigr)^2}
=\frac{2|\Delta|}{\pi\kappa}=\frac{2Q}{\pi}.
\label{eq:IPR}
\end{equation}
The coupling $g$ cancels between numerator and denominator, and the prefactor $2|\Delta|$ makes the ratio dimensionless. Because that prefactor is a chosen scale, $\mathrm{IPR}_{\Delta}$ measures spectral width relative to the detuning and is equivalent to $Q$ only for the Lorentzian family used here.
Rewriting the high-$Q$ expansion~\eqref{eq:highQ} in terms of $\mathrm{IPR}_{\Delta}$ gives
\begin{equation}
\mathcal C_g=1-\frac{1}{4\,\mathrm{IPR}_{\Delta}^{\,2}}+\mathcal O(\mathrm{IPR}_{\Delta}^{-3}),\qquad Q\gg1.
\label{eq:IPR_relation}
\end{equation}
The scaling $\mathcal C_g-1\propto-1/\mathrm{IPR}_{\Delta}^2$ provides a formal comparison with participation measures used in localization theory~\cite{Mirlin2000,Lee1985,Kramer1993}. Here, $|J|/\Gamma=Q$ measures coherent exchange relative to collective decay. In localization theory, participation ratios characterize the spatial concentration of wave functions and its relation to transport. The analogy lies in this common mathematical description of concentration, while the underlying mechanisms remain different.
The narrow- and broad-linewidth limits therefore characterize the response of this cavity-mediated protocol.

The predicted $\mathcal C_g(Q)$ curve is, in principle, directly observable in superconducting circuit QED experiments~\cite{Clarke2008,Devoret2013,Kjaergaard2020}. One prepares the $|+\rangle|+\rangle$ state, waits until $t_g=\pi/(2|J|)$, post-selects the no-jump trajectory, and extracts the concurrence by quantum state tomography. Comparison with Eq.~\eqref{eq:Cg} assumes efficient monitoring and negligible additional qubit relaxation or dephasing. The high- and low-$Q$ asymptotics then provide tests of the conditional cavity model.

\section{Conclusion and Outlook}
\label{sec:conclusion}

In this work, we constructed a minimal, analytically solvable model of two qubits coupled through a lossy single-mode cavity and derived the closed-form no-jump concurrence at the reference gate time
\begin{equation}
\mathcal C_g(Q)=\frac{2e^{-\pi/(2Q)}\bigl(1+e^{-\pi/(2Q)}\bigr)}{1+3e^{-\pi/Q}},\quad
Q\equiv\frac{|J|}{\Gamma}=\frac{|\Delta|}{\kappa}.
\end{equation}
which depends solely on the detuning-to-linewidth ratio $Q$, proportional to the detuning-normalized participation ratio $\mathrm{IPR}_{\Delta}$ for the Lorentzian response. It interpolates analytically between the weak-loss and strongly overdamped limits of the specified conditional protocol. Accordingly, $\mathcal C_g\simeq1-\pi^{2}/(16Q^{2})$ at high $Q$, whereas at low $Q$ it decays exponentially, $\mathcal C_g\simeq2e^{-\pi/(2Q)}$. These limits give the spectral dependence for the fixed initial state, no-jump record, and reference time considered here.

The Reeh--Schlieder theorem provides broader field-theoretic background for the nontrivial correlation structure of the quantum vacuum. Within this broader context, the present result characterizes how the spectral response of a cavity-mediated conditional channel controls two-qubit entanglement. A direct quantitative comparison with correlations between spacelike regions would require a spatially resolved detector--field model.

The relation $\mathcal C_g-1\propto-1/\mathrm{IPR}_{\Delta}^2$ uses a participation measure familiar from Anderson localization~\cite{Anderson1958,Evers2008}. In both settings, a participation ratio provides a compact description of concentration. Here it characterizes a Lorentzian spectral response and the balance between coherent exchange and collective decay, whereas in Anderson localization it characterizes wave-function structure and transport.

The predicted $\mathcal C_g(Q)$ curve is, in principle, directly observable in superconducting circuit QED experiments, where the concurrence can be extracted via no-jump post-selected quantum state tomography. The parameter-free asymptotic coefficients provide sharp, falsifiable predictions for this conditional model. Vacuum-correlation-extraction protocols in superconducting circuits have been analyzed theoretically~\cite{Sabin2012}, and the on-chip platform also enables the quantum simulation of open-system dynamics~\cite{Houck2012}. Together, these capabilities make the platform suitable for comparing the present cavity-mediated protocol with vacuum-correlation-extraction protocols.

Several avenues warrant further exploration. First, extending the spectral-localization analysis to non-Lorentzian spectral densities, such as those of photonic crystal cavities~\cite{John1987,Yablonovitch1987} or multimode waveguide QED platforms~\cite{Houck2012}, would test which features survive beyond the Lorentzian family. This is most pertinent near $Q\sim1$, where the Markovian approximation begins to fail and non-Markovian memory effects or information backflow may appear~\cite{Breuer2016,Rivas2014}. Second, generalizing to $N$-qubit arrays would enable the study of collective superradiant and subradiant dynamics in entanglement routing; just as the antisymmetric Bell state $|\Psi_-\rangle$ is dark under the collective decay in Sec.~\ref{sec:concurrence}, subradiant states of $N$-qubit arrays are immune to this decay channel, allowing entanglement to be stored without dissipation~\cite{Francica2009}. Third, treating entanglement as a fuel extracted from a quantum field reservoir may yield a thermodynamic interpretation of spectral localization~\cite{Vinjanampathy2016}. Finally, a spatially resolved detector-field extension could connect the spectral response studied here with the accessibility of vacuum correlations in curved spacetimes and interacting quantum fields.

\begin{acknowledgments}
Hao Xu thanks National Natural Science Foundation of China (No.12205250) for funding support.
\end{acknowledgments}

\end{document}